# X-ray Photoemission Study of the Infinite-Layer Cuprate Superconductor $Sr_{0.9}La_{0.1}CuO_2$

**R. P. Vasquez[a]\*, C. U. Jung[b], J. Y. Kim[b], Min-Seok Park[b], Heon-Jung Kim[b], and Sung-Ik Lee[b]**

[a]**Center for Space Microelectronics Technology, Jet Propulsion Laboratory, California Institute of Technology, Pasadena, California 91109-8099, USA**

[b]**National Creative Research Initiative Center for Superconductivity and Department of Physics, Pohang University of Science and Technology, Pohang 790-784, Republic of Korea**

**Short Title: XPS of Infinite-Layer Cuprate**

\*email: richard.p.vasquez@jpl.nasa.gov

**Abstract**

The electron-doped infinite-layer superconductor $Sr_{0.9}La_{0.1}CuO_2$ is studied with x-ray photoemission spectroscopy (XPS). A nonaqueous chemical etchant is shown to effectively remove contaminants and to yield surfaces from which signals intrinsic to the superconductor dominate. These data are compared to measurements from hole-doped $La_{1.85}Sr_{0.15}CuO_4$, from undoped $La_2CuO_4$, and from electron-doped $Nd_{1.85}Ce_{0.15}CuO_{4-\delta}$. The Cu 2p core level is consistent with a lower value of the O 2p $\rightarrow$ Cu 3d charge transfer energy $\Delta$ than in hole-doped cuprates. A clear Fermi edge is observed in the valence band region.

## 1. Introduction

High temperature superconductors based on $SrCuO_2$ have $CuO_2$ planes which are separated only by $Sr^{2+}$ ions (e.g. see [1]), without the intervening charge reservoir layers present in other families of cuprate superconductors, and hence are referred to as infinite-layer systems. Very recent scanning tunneling spectroscopy measurements [2] have determined that $Sr_{0.9}La_{0.1}CuO_2$ exhibits several unique features not observed in other cuprate superconductors, including isotropic s-wave pairing symmetry, insignificant spin fluctuations, and no pseudogap observed above the superconducting transition temperature $T_c$. Studies of this material are thus of fundamental interest. $SrCuO_2$ is a charge transfer insulator in which the Cu-O layers consist of zigzag chains when synthesized at ambient pressure. The infinite-layer structure with $CuO_2$ planes is attained by high pressure synthesis and superconductivity is achieved with either Sr vacancies (hole doping) or partial substitution of rare earth ions such as $Nd^{3+}$ or $La^{3+}$ on the $Sr^{2+}$ site (electron doping).

Photoemission has proven to be a powerful technique in studies of fundamental physical properties of cuprate superconductors. While much recent photoemission work has been focused on information such as Fermi surface mapping obtained from high resolution angle-resolved photoemission near the Fermi level, core level measurements can also yield valuable information on trends in chemical bonding correlated with physical properties of these materials (e.g. see [3]). Previous photoemission studies of $SrCuO_2$-based materials [4-10], including electron-doped $Sr_{0.85}Nd_{0.15}CuO_2$ [4] and hole-doped $(Sr_{1-x}Ca_x)_{0.9}CuO_2$ [6], have primarily involved valence band measurements, with only a single study of undoped $SrCuO_2$ [10] on the Cu 2p core levels. In this work, x-ray photoemission spectroscopy (XPS) is used to study the core levels and valence band of the electron-doped infinite-layer superconductor $Sr_{0.9}La_{0.1}CuO_2$. These data are compared to measurements from the hole-doped superconductor $La_{1.85}Sr_{0.15}CuO_4$, which has the same chemical constituents as $Sr_{0.9}La_{0.1}CuO_2$, from the undoped parent compound $La_2CuO_4$ (which is an antiferromagnetic insulator), and from the electron-doped superconductor $Nd_{1.85}Ce_{0.15}CuO_{4-\delta}$.

## 2. Experimental

A polycrystalline pellet of $Sr_{0.9}La_{0.1}CuO_2$ with $T_c = 43$ K is synthesized at high pressure and characterized as described elsewhere [11]. An SEM micrograph of a pellet is shown in Figure 1. 3000 Å-thick epitaxial films of $La_2CuO_4$ and $La_{1.85}Sr_{0.15}CuO_4$ are grown by pulsed laser deposition in 100 mTorr background pressure of oxygen onto $LaAlO_3$ (100) substrates at 750° C. The as-grown $La_{1.85}Sr_{0.15}CuO_4$ has $T_c = 13$ K, as determined by ac susceptibility measurement, with a transition width $\Delta T_c = 1$ K, increasing to $T_c = 26$ K with $\Delta T_c = 2$ K after annealing in 1 atmosphere oxygen at 800° C for 8 hours. Data for $Nd_{1.85}Ce_{0.15}CuO_{4-\delta}$ epitaxial films have been previously reported [12].

The sample surfaces are cleaned in a dry box with an inert ultrahigh purity $N_2$ atmosphere which encloses the load lock area of the XPS spectrometer. A nonaqueous etchant consisting of 0.5% $Br_2$ in absolute ethanol is used for the polycrystalline pellet, and 0.2% HCl in absolute ethanol is used for the epitaxial films. The effectiveness of nonaqueous etching in producing clean surfaces of high temperature superconductors has been previously demonstrated [13,14]. The samples are immersed in the etchant for 15 seconds, rinsed in ethanol, blown dry with nitrogen, and loaded into the XPS spectrometer with no atmospheric exposure. The measured surface stoichiometry of the $Sr_{0.9}La_{0.1}CuO_2$ is Sr:La:Cu:O = 1.01:0.14:0.85:2.27. The slight deviation from the expected bulk stoichiometry results from residual surface and/or grain boundary contaminants, evident in the XPS spectra as residual high binding energy signals in the O 1s and Sr 3d core level regions.

The XPS spectrometer is a Surface Science Instruments SSX-501 utilizing monochromatic Al Kα x-rays (1486.6 eV). Spectra are measured at ambient temperature with photoemission 55° from the surface normal. The spectra are measured with an x-ray spot size of 150 or 300 μm and the pass energy of the electron energy analyzer is 25 eV. The energy scale is calibrated using sputter-cleaned Au and Cu with the Au $4f_{7/2}$ binding energy set to 83.95 ± 0.05 eV (0.85 eV full width at half-maximum (FWHM)) and the Cu $2p_{3/2}$ binding energy set to 932.45 ± 0.05 eV (1.0 eV FWHM).

### 3. Results and Discussion

The results of the XPS measurements are summarized in Table 1. The spectra presented in the figures have been normalized to the same peak height to facilitate comparison of peak positions and shapes. The O 1s core levels measured after surface cleaning are shown in Figure 2. The signals consist of a low binding energy signal near 528-529 eV corresponding to the cuprates, and a high binding energy signal near 531-532 eV corresponding to residual surface contaminants. Prior to etching, only the contaminant signal is detectable from the $Sr_{0.9}La_{0.1}CuO_2$ surface. After etching, the cuprate signal is dominant but the contaminant signal remains more prominent compared to measurements from the other materials, probably due to the higher Sr content in $Sr_{0.9}La_{0.1}CuO_2$ since alkaline earth components in high temperature superconductors react readily with air to form carbonates [14]. Attempts to obtain cleaner surfaces by additional etching were unsuccessful, resulting in insulating surfaces which exhibited charging-induced signal shifts. This observation most likely results from the etching of grain boundaries in the polycrystalline sample, yielding poor electrical contact between grains.

The Sr 3d core levels are shown in Figure 3. The Sr 3d signal from $La_{1.85}Sr_{0.15}CuO_4$ in Figure 3(a) is situated on top of a broad energy loss peak from the La 4d signal, which is much lower intensity in the Sr 3d spectrum from $Sr_{0.9}La_{0.1}CuO_2$ shown in Figure 3(b). Both of the spectra in Figure 3 are well-represented by two spin-orbit split doublets with the expected 3:2 ratio. The higher binding energy doublet has a $3d_{5/2}$ component near 133 eV which is dominant prior to etching but is a minor component after etching, and is consistent with $SrCO_3$ [15]. The lower binding energy doublet is dominant in each spectrum after etching and has a $3d_{5/2}$ component with a binding energy near 131.8 eV originating from the superconducting phase. The observed binding energies from the superconducting phases are consistent with measurements from other Sr-containing cuprate superconductors [15,16], and are significantly lower than the Sr $3d_{5/2}$ binding energy measured from Sr metal [15]. Simpler alkaline earth compounds also exhibit this unusual negative chemical shift [15], which has been attributed to initial-state electrostatic effects primarily due to the large value of the Madelung energy relative to the ionization energy [15-18].

The La 3d core levels are shown in Figure 4. The low signal-to-noise ratio in the spectrum measured from $Sr_{0.9}La_{0.1}CuO_2$ shown in Figure 4(c) is due to the weak signal since the La content is low. Each spin-orbit split component has a complex double peak structure typical of $La^{3+}$ compounds [19,20] due to $\underline{3d}^9 4f^0 L$ and $\underline{3d}^9 4f^1 \underline{L}$ state mixing [21], where L denotes the oxygen ligand and underscoring denotes a hole. In previous measurements in this lab [22], the intensity ratio and energy separation of the two peaks in each spin-orbit split component were found to be the same for different perovskite and layered cuprate materials and insensitive to differences in doping, consistent with the spectra in Figure 4. Note that all of the core level spectra measured from hole-doped $La_{1.85}Sr_{0.15}CuO_4$ are rigidly shifted to lower binding energy by approximately 0.2 eV relative to those measured from undoped $La_2CuO_4$ (see Table 1 and curves (a) and (b) in Figures 2, 4, and 5). This observation is consistent with the doping-induced shift of the chemical potential, which has been previously reported for $La_{2-x}Sr_xCuO_4$ [23,24], as well as for other cuprate superconductors [25-27].

The Cu $2p_{3/2}$ core level spectra are presented in Figure 5. The multiplet at higher binding energy, referred to as a satellite peak in the literature, corresponds to states of predominant $\underline{2p}^5 3d^9 L$ character, while the main peak at lower binding energy corresponds to states of predominant $\underline{2p}^5 3d^{10} \underline{L}$ character resulting from ligand-to-metal (O $2p \rightarrow$ Cu $3d$) charge transfer [28-30]. Different final states can have sufficient energy separation to yield resolvable features in the main peak, which is most evident in the spectrum measured from $Sr_{0.9}La_{0.1}CuO_2$ in Figure 5(c). Such features have previously been resolved in Cu 2p spectra measured from $SrCuO_2$ [10], in which the Cu-O layers consist of zigzag chains, and from $Sr_2CuO_3$ [10,31], in which the Cu-O layers consist of linear chains. Calculations for $Sr_2CuO_3$ based on a three-band Hubbard model have also been shown to be consistent with the measured Cu 2p spectrum [31], with the main peak features corresponding to final states with different screening of the photoexcited core hole.

The Cu 2p spectra are commonly analyzed within a simple configuration interaction model utilizing a two-band Hamiltonian [3,29,30,32,33]. Within this model, the O 2p → Cu 3d charge transfer energy $\Delta$, the O 2p - Cu 3d hybridization strength T, and the on-site Coulomb interaction between Cu 2p and Cu 3d holes U, are related to the experimentally-determined energy separation between the poorly-screened satellite and well-screened main peak ($E_s$ - $E_m$) and to the ratio of the intensities of the satellite and main peak ($I_s/I_m$). The experimental values of $I_s/I_m$ and $E_s$ - $E_m$ obtained for $La_{1.85}Sr_{0.15}CuO_4$ are consistent with the earlier results for hole-doped cuprates [3]. The electron-doped materials both exhibit higher values of $E_s$ - $E_m$ and lower values of $I_s/I_m$ than values measured from hole-doped cuprates. One might have expected that the Cu 2p spectra of $Nd_{1.85}Ce_{0.15}CuO_{4-\delta}$ and $Sr_{0.9}La_{0.1}CuO_2$ would be most similar. Both materials are electron doped, both have similar Cu-O bond lengths [1,34] (1.97 Å) which are significantly greater than those in $La_{2-x}Sr_xCuO_4$ [35] (1.89 Å) and the coordination of O to Cu is square planar in both materials whereas it is distorted octahedral in $La_{2-x}Sr_xCuO_4$. However, the values of $I_s/I_m$ and $E_s$ - $E_m$ obtained for $Sr_{0.9}La_{0.1}CuO_2$ are closer to those measured from undoped cuprates [3]. The higher values of $E_s$ - $E_m$ and lower values of $I_s/I_m$ observed are consistent with undoped and electron-doped cuprates having higher values of T or lower values of $\Delta$ compared to the hole-doped cuprates. Since the Cu-O bond lengths in $Nd_{1.85}Ce_{0.15}CuO_{4-\delta}$ and $Sr_{0.9}La_{0.1}CuO_2$ are longer than in La$_{2-x}$Sr$_x$CuO$_4$, a higher degree of O 2p – Cu 3d hybridization is unlikely for these materials. The Cu 2p spectra are thus most consistent with the electron-doped materials having lower values of $\Delta$ compared to hole-doped cuprates.

The valence band spectra are presented in Figure 6. The valence bands consist of hybridized Cu 3d – O 2p states, for the photon energy used here the Cu 3d states dominate due to the difference in the photoionization cross sections [36]. For Nd$_{1.85}$Ce$_{0.15}$CuO$_{4-\delta}$, Nd 4f states also have a non-negligible contribution to the measured spectrum, particularly on the high binding energy side. The observation of a clear Fermi edge in the Sr$_{0.9}$La$_{0.1}$CuO$_2$ spectrum, similar to results obtained on other chemically etched cuprate superconductor surfaces, is further evidence that the nonaqueous chemical etching effectively removes surface contaminants.

## 4. Summary and Conclusions

The core levels and valence band of the electron-doped infinite-layer superconductor Sr$_{0.9}$La$_{0.1}$CuO$_2$ have been measured with XPS. A nonaqueous chemical etchant has been shown to effectively remove surface hydroxides and carbonates which result from air exposure and to yield surfaces from which signals intrinsic to the superconducting phase dominate. These data have been compared to measurements from the hole-doped superconductor La$_{1.85}$Sr$_{0.15}$CuO$_4$, which has the same chemical constituents as Sr$_{0.9}$La$_{0.1}$CuO$_2$, from undoped La$_2$CuO$_4$, and from Nd$_{1.85}$Ce$_{0.15}$CuO$_{4-\delta}$, which is also electron-doped and has similar Cu-O bond lengths and the same square planar O coordination to Cu as Sr$_{0.9}$La$_{0.1}$CuO$_2$. The O 1s, Sr 3d, and La 3d core levels of Sr$_{0.9}$La$_{0.1}$CuO$_2$ are similar to those measured from hole-doped cuprates. The Cu 2p core level is consistent with a lower value of the O 2p $\rightarrow$ Cu 3d charge transfer energy $\Delta$ than in hole-doped cuprates. A clear Fermi edge is observed in the valence band region, further evidence of the effectiveness of the chemical etching.

**Acknowledgments**

Part of the work described in this paper was performed by the Center for Space Microelectronics Technology, Jet Propulsion Laboratory, California Institute of Technology, and was sponsored by the National Aeronautics and Space Administration. Part of this work was performed at Pohang University of Science and Technology, and supported by the Ministry of Science and Technology of Korea through the Creative Research Initiative Program.

**Table 1.** Summary of core level binding energies (±0.05 eV) and peak full widths at half maximum (in parentheses) for the materials considered in this work. The signal centroids are used for the Cu $2p_{3/2}$ main peak binding energies and satellite - main peak energy separations ($E_s$-$E_m$), the satellite - main peak intensity ratios ($I_s/I_m$) are also listed.

| **Material** | **Sr $3d_{5/2}$** | **La $3d_{5/2}$** | **Cu $2p_{3/2}$** | **$I_s/I_m$** | **$E_s$-$E_m$** | **O 1s** | **Source** |
|---|---|---|---|---|---|---|---|
| $Sr_{0.9}La_{0.1}CuO_2$ | 131.75 (1.25) | 833.8 | 933.5 (3.9) | 0.35 | 8.8 | 528.5 (1.25) | This work |
| $Nd_{1.85}Ce_{0.15}CuO_{4-\delta}$ | none | none | 933.65 (4.2) | 0.26 | 9.6 | 528.9 (1.25) | [11] |
| $La_2CuO_4$ | none | 833.15 | 933.5 (2.95) | 0.41 | 8.75 | 528.5 (1.35) | This work and [2] |
| $La_{1.85}Sr_{0.15}CuO_4$ | 131.8 (1.05) | 832.9 | 933.3 (3.25) | 0.38 | 8.6 | 528.3 (1.40) | This work |

## Figure Captions

**Figure 1.**   SEM micrograph of a $Sr_{0.9}La_{0.1}CuO_2$ pellet.

**Figure 2.**   O 1s core levels measured from chemically-etched surfaces of (a) $La_{1.85}Sr_{0.15}CuO_4$ epitaxial film, (b) $La_2CuO_4$ epitaxial film, (c) $Sr_{0.9}La_{0.1}CuO_2$ polycrystalline pellet, and (d) $Nd_{1.85}Ce_{0.15}CuO_4$ epitaxial film.

**Figure 3.**   Sr 3d core levels measured from chemically-etched surfaces of (a) $La_{1.85}Sr_{0.15}CuO_4$ epitaxial film, and (b) $Sr_{0.9}La_{0.1}CuO_2$ polycrystalline pellet.

**Figure 4.**   La 3d core levels measured from chemically-etched surfaces of (a) $La_{1.85}Sr_{0.15}CuO_4$ epitaxial film, (b) $La_2CuO_4$ epitaxial film, and (c) $Sr_{0.9}La_{0.1}CuO_2$ polycrystalline pellet.

**Figure 5.**   Cu $2p_{3/2}$ core levels measured from chemically-etched surfaces of (a) $La_{1.85}Sr_{0.15}CuO_4$ epitaxial film, (b) $La_2CuO_4$ epitaxial film, (c) $Sr_{0.9}La_{0.1}CuO_2$ polycrystalline pellet, and (d) $Nd_{1.85}Ce_{0.15}CuO_4$ epitaxial film.

**Figure 6.**   Valence bands measured from chemically-etched surfaces of (a) $La_{1.85}Sr_{0.15}CuO_4$ epitaxial film, (b) $La_2CuO_4$ epitaxial film, (c) $Sr_{0.9}La_{0.1}CuO_2$ polycrystalline pellet, and (d) $Nd_{1.85}Ce_{0.15}CuO_4$ epitaxial film.  Inset: Fermi level region.

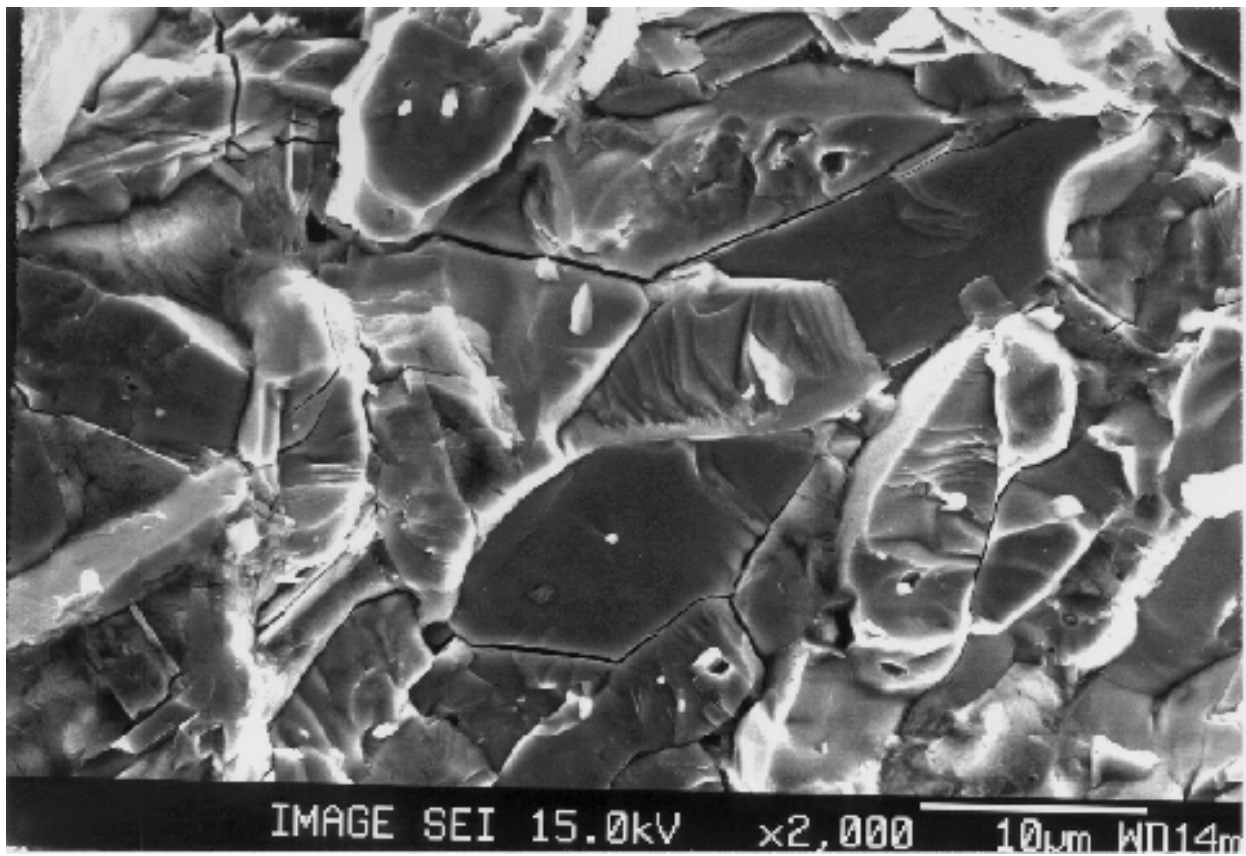

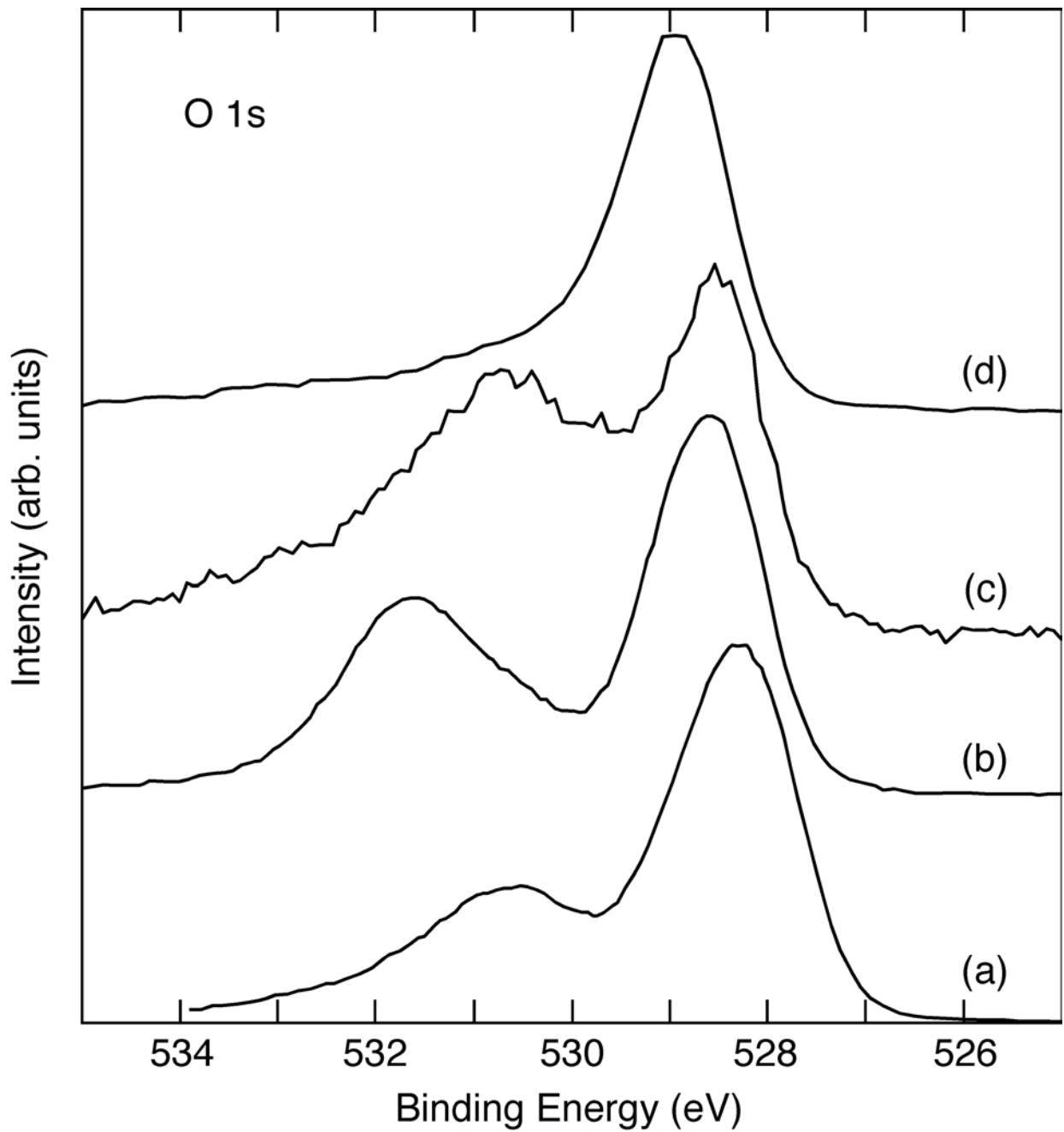

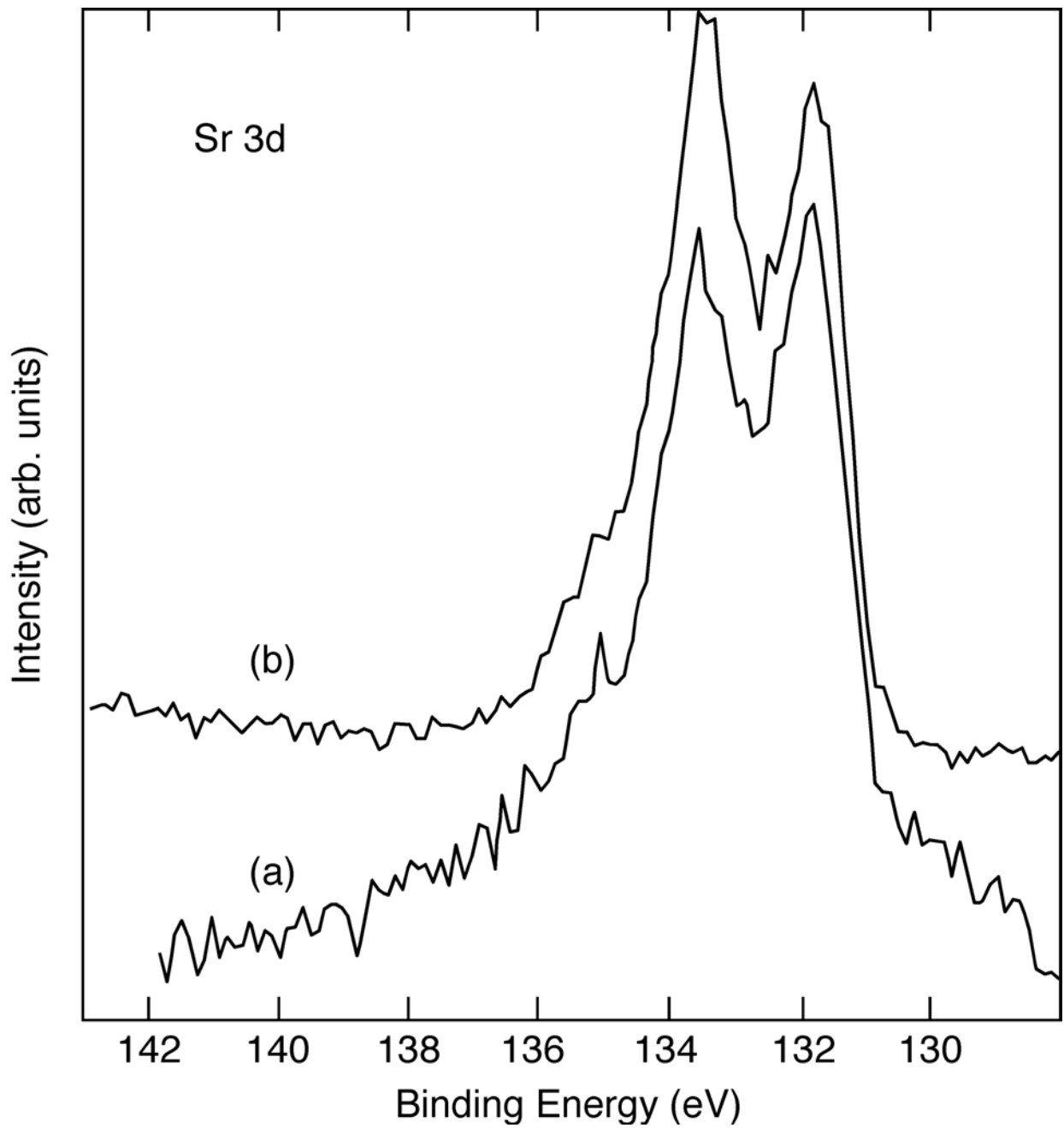

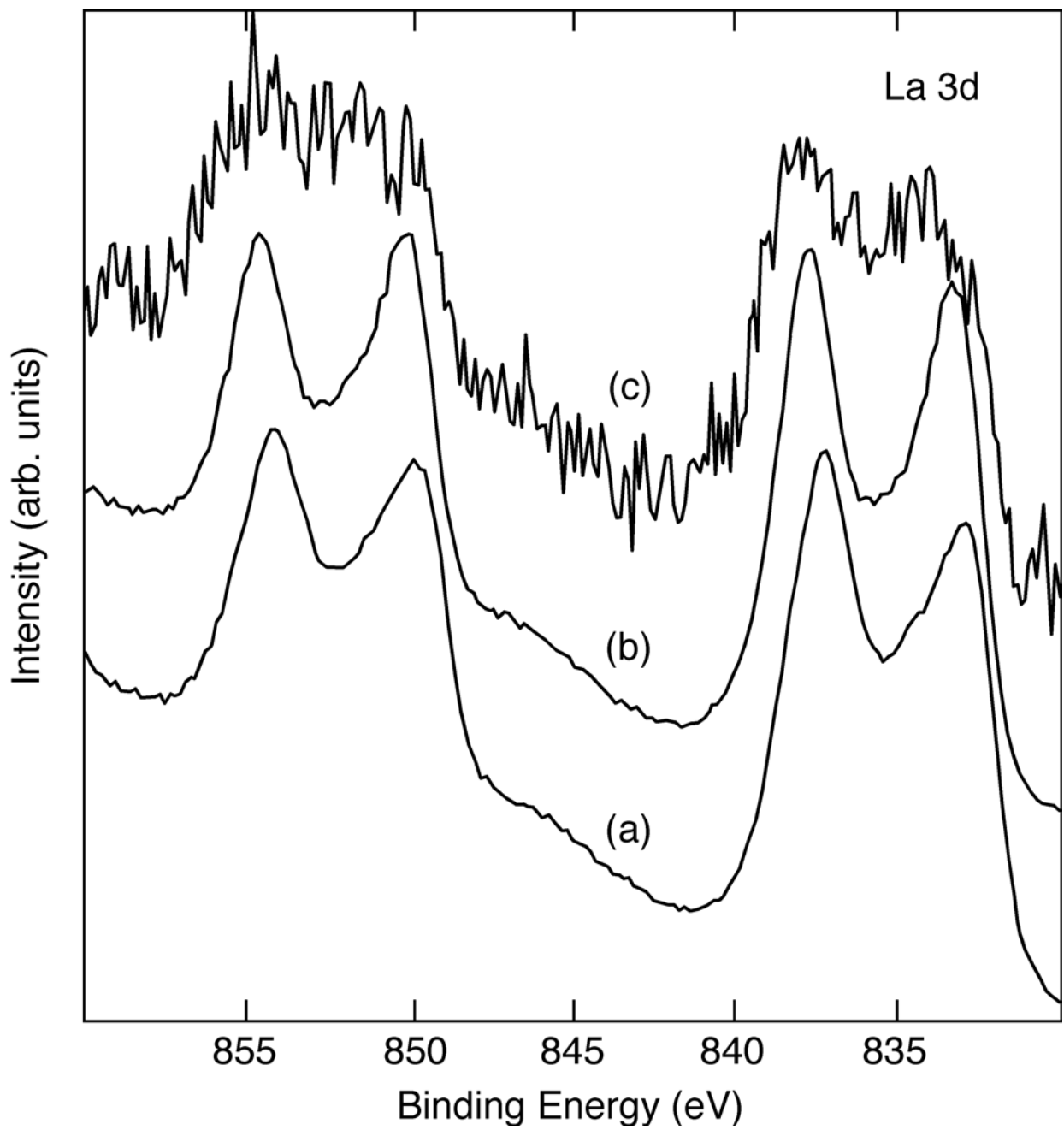

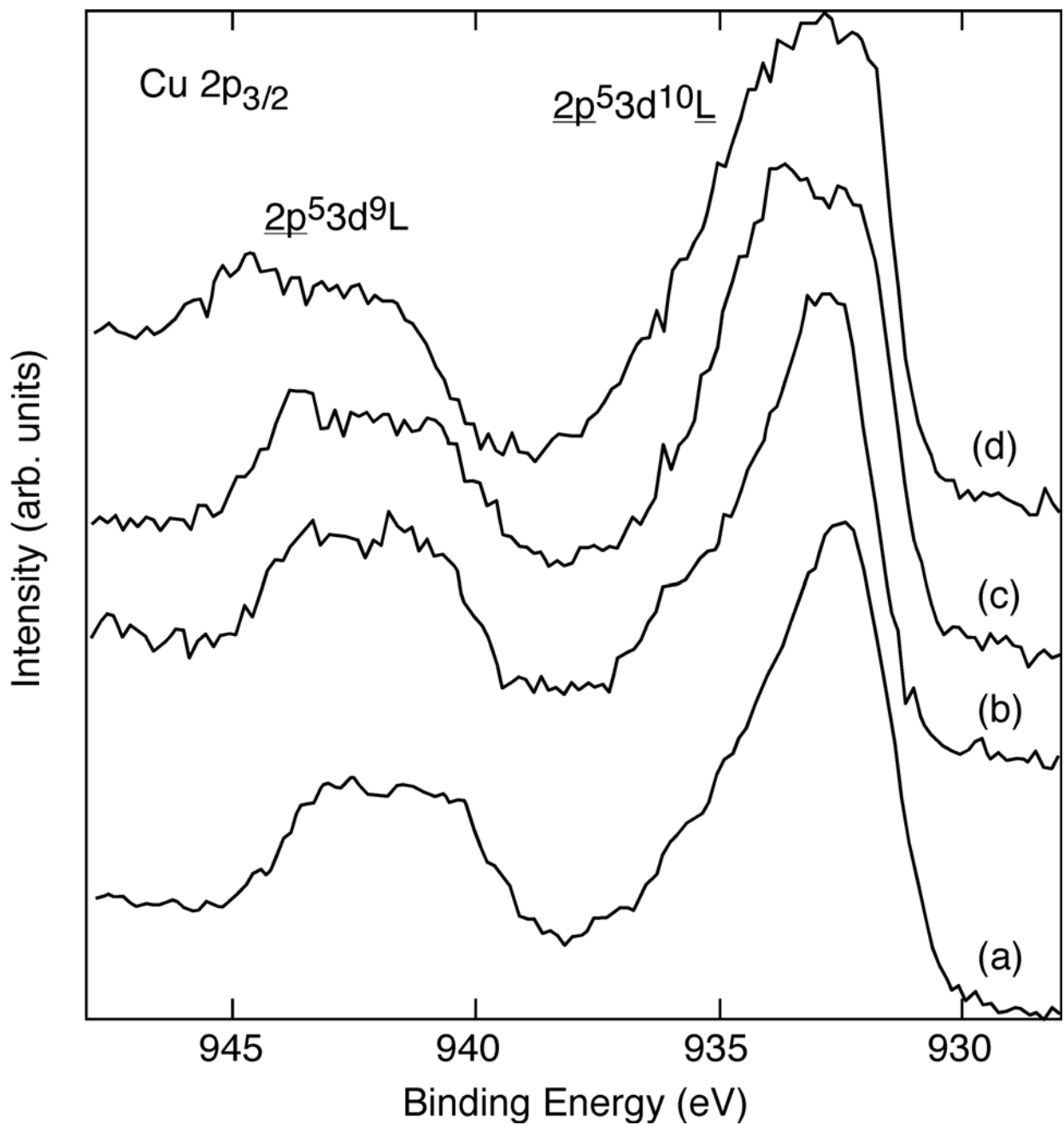

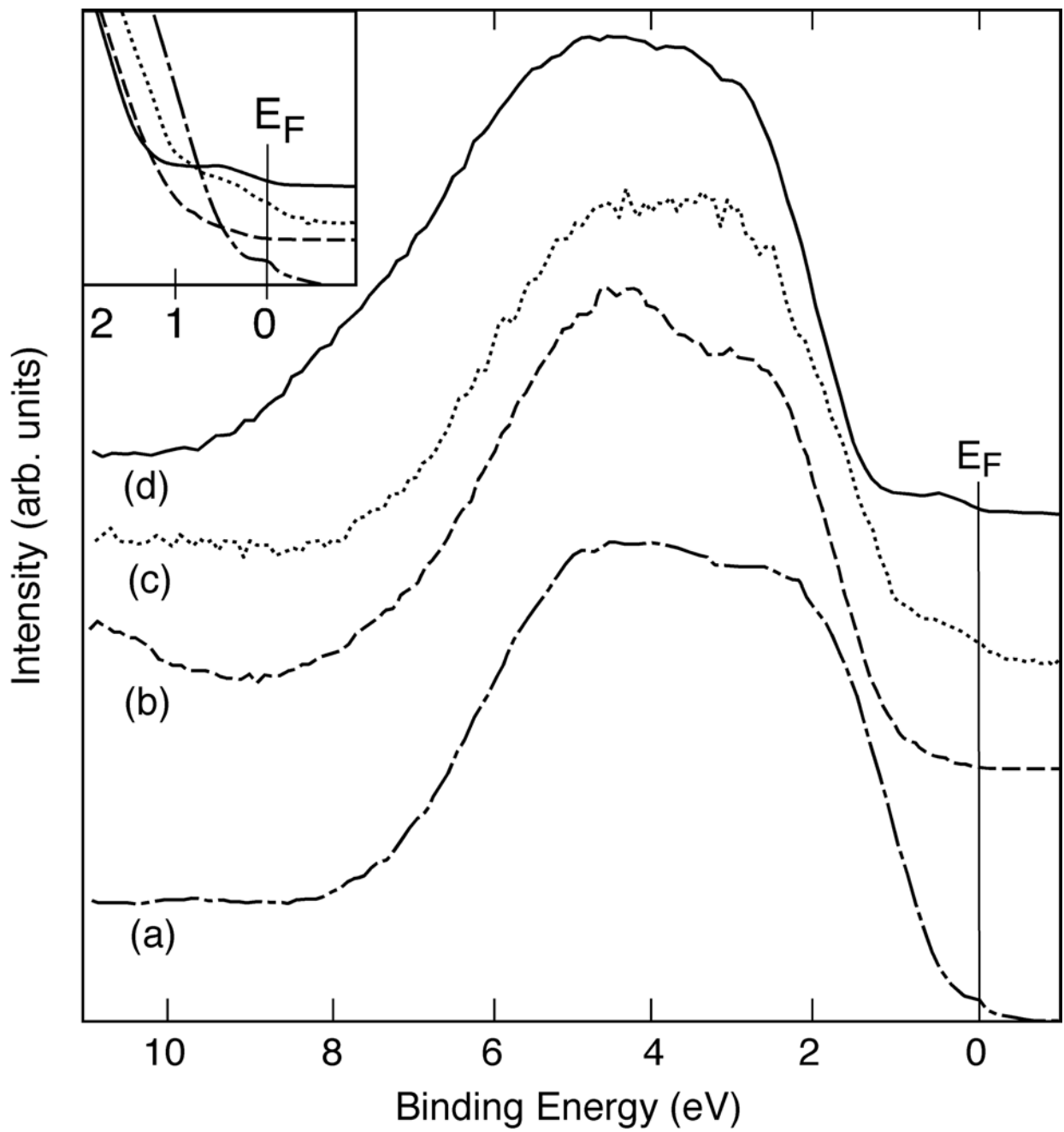